\documentclass{article}
\usepackage{spconf}
\usepackage{cite}
\usepackage{graphicx}
\usepackage{subfigure}
\usepackage{amsfonts}
\usepackage{lipsum}
\usepackage{float}
\usepackage{amsmath}
\usepackage{nicefrac}
\usepackage{xcolor}
\usepackage{hyperref}
 \usepackage{multirow}
\usepackage{algorithm}
\usepackage{diagbox}
\usepackage[noend]{algpseudocode}
\usepackage{arydshln}

\DeclareMathAlphabet{\mathcal}{OMS}{cmsy}{m}{n}
\usepackage{enumitem}

\usepackage{url}

\usepackage{hyperref} 

\begin{document}
	
	\onecolumn
	
	\begin{description}[labelindent=0cm,leftmargin=3cm,rightmargin=3cm,style=multiline]
		
		\item[\textbf{Citation}]{M. Alfarraj and G. AlRegib, "Semi-supervised Learning for Acoustic Impedance Inversion," SEG Technical Program Expanded Abstracts 2019. Society of Exploration Geophysicists}

		\item[\textbf{Review}]{Accepted on: 23 May 2019}
		
		\item[\textbf{Data and Codes}]{\href{https://github.com/olivesgatech/Semi-supervised-Learning-for-Acoustic-Impedance-Inversion}{[\underline{GitHub Link}]}}% If you do not have data related to this paper, you can remove the data keyword.

		\item[\textbf{Bib}] {@incollection\{alfarraj2019semisupervised,\\
  title={Semi-supervised Learning for Acoustic Impedance Inversion},\\
  author={Alfarraj, Motaz and AlRegib, Ghassan},\\
  booktitle={SEG Technical Program Expanded Abstracts},\\
  year={2019},\\
  publisher={Society of Exploration Geophysicists\}\\
}}

%		\item[\textbf{Copyright}]{\textcopyright 2018 IEEE. Personal use of this material is permitted. Permission from IEEE must be obtained for all other uses, in any current or future media, including reprinting/republishing this material for advertising or promotional purposes,
%			creating new collective works, for resale or redistribution to servers or lists, or reuse of any copyrighted component
%			of this work in other works. }
		
		\item[\textbf{Contact}]{\href{mailto:motaz@gatech.edu}{motaz@gatech.edu}  OR \href{mailto:alregib@gatech.edu}{alregib@gatech.edu}\\ \url{http://ghassanalregib.com/} \\ }
	\end{description}
	
	%Following command sequence was used to start the paper content from the following page and avoid numbering cover page.
	\thispagestyle{empty}
	\newpage
	\clearpage
	\setcounter{page}{1}
	
	%Cover page was 1 column. \twocolumn changes the page format back to double column.
	\twocolumn
	
\title{Semi-supervised Learning for Acoustic Impedance Inversion}
\name{Motaz Alfarraj and Ghassan AlRegib}
\address{Center for Energy and Geo Processing (CeGP) \\ School of Electrical and Computer Engineering \\ Georgia Institute of Technology \\ \{motaz,alregib\}@gatech.edu}

\maketitle

\begin{abstract}
Recent applications of deep learning in the seismic domain have shown great potential in different areas such as inversion and interpretation. Deep learning algorithms, in general, require tremendous amounts of labeled data to train properly. To overcome this issue, we propose a \emph{semi-supervised framework for acoustic impedance inversion based on convolutional and recurrent neural networks}. Specifically, seismic traces and acoustic impedance traces are modeled as time series. Then, a neural-network-based inversion model comprising convolutional and recurrent neural layers is used to invert seismic data for acoustic impedance. The proposed workflow uses well log data to guide the inversion. In addition, it utilizes a learned seismic forward model to regularize the training and to serve as a geophysical constraint for the inversion. The proposed workflow achieves an average correlation of $98\%$ between the estimated and target elastic impedance using 20 AI traces for training. 
\end{abstract}

\section{Introduction}
Seismic inversion is the process of estimating rock properties from seismic reflection data. In principle, inversion is a procedure to infer true model parameters $m \in X$ through indirect measurements $d \in Y$. Mathematically, the problem can be formulated as follows
\begin{equation}
    d = \mathcal{F}(m) + n,
    \label{eqn:system_setup}
\end{equation}

where $\mathcal{F}: X\rightarrow Y$ is a forward operator, $d$ is the measured data, $m$ is the model, and $n\in Y$ is a random variable that represents noise in the measurements. To estimate the model from the measured data, one needs to solve an inverse problem. The solution depends on the nature of the forward model and observed data. In the case of seismic inversion, and due to the non-linearity and heterogeneity of the subsurface, the inverse problem is ill-posed. In order to find a stable solution to an ill-posed problem, the problem needs to be regularized. For instance, one can seek a solution by imposing constraints on the solution space, or by incorporating prior knowledge about the model. 
A classical approach to solve inverse problems is to set up the problem as a Bayesian inference problem, and improve prior knowledge by optimizing for a cost function based on the data likelihood,
\begin{equation}
    \hat{m} = \underset{m\in X}{\text{argmin}} \left[ \mathcal{H}\left(\mathcal{F}(m),d\right)+ \lambda \mathcal{C}(m)\right],
    \label{eqn:classical_inversion}
\end{equation} 

where $\hat{m}$ is the estimated model, $\mathcal{H}:Y\times Y \rightarrow \mathbb{R}$ is an affine transform of the data likelihood, $\mathcal{C}:X \rightarrow \mathbb{R}$ is a regularization function that incorporates prior knowledge in the inversion, and $\lambda$ is regularization parameters that control the influence of the regularization function. 

The solution of equation \ref{eqn:classical_inversion} in seismic inversion can be sought in a stochastic or a deterministic fashion through an optimization routine. The literature of seismic inversion is rich in various methods to formulate, regularize and solve the problem (e.g., \cite[]{duijndam1988bayesian_1, doyen1988porosity, duijndam1988bayesian_2, ulrych2001bayes, buland2003bayesian, gholami2015nonlinear, tarantola2005inverse}).

Recently, there have been several successful applications of machine learning and deep learning methods in inverse problems \cite[]{lucas2018using}. Moreover, machine learning and deep learning methods have been utilized in the seismic domain for different tasks such as inversion and interpretation \cite[]{alregib2018subsurface}. For example, seismic inversion has been attempted using supervised-learning algorithms such as support vector regression (SVR) \cite[]{al2012support}, artificial neural networks \cite[]{roth1994neural,araya2018deep}, committee models \cite[]{gholami2017estimation}, convolutional neural networks (CNNs) \cite[]{das2018convolutional}, recurrent neural networks \cite[]{alfarraj2018petrophysical}, and many other methods \cite[]{chaki2015novel,yuan2013spectral, gholami2017estimation,chaki2017diffusion,mosser2018rapid,chaki2018well}. 

In general, machine learning algorithms are used to learn a non-linear mapping parameterized by $\Theta\in Z$, i.e., $\mathcal{F}_{\Theta}^\dagger: Y\rightarrow X$ from a set of examples (known as the training dataset) such that:
\begin{equation}
    \mathcal{F}_{\Theta}^\dagger(d) \approx m.
    \label{eqn:machine_learning_inversion}
\end{equation}

There is one key difference between classical inversion methods and machine learning methods. In classical inversion, the outcome is a set of model parameters (deterministic) or a posterior probability density function (stochastic). On the other hand, learning methods produce a mapping from measurements domain to model parameters domain ($\mathcal{F}^{\dagger}_\Theta$). 

Using neural networks, one can learn $\mathcal{F}^{\dagger}_\Theta$ (in equation \ref{eqn:machine_learning_inversion}) using different learning schemes such as supervised or unsupervised learning \cite[]{adler2017solving}. In supervised learning, the machine learning algorithm is given a set measurement-model pairs $\{d,m\}$ (e.g., seismic traces and their corresponding rock property traces from well logs) to learn the mapping by minimizing the following loss function
\begin{equation}
    L(\Theta):=\mathcal{D}\left(\hat{m}, \mathcal{F}_{\Theta}^\dagger(d)\right)
    \label{eqn:supervised}
\end{equation}

where $\mathcal{D}$ is a distance measure that compares the estimated rock property to the estimated property. Namely, supervised machine learning algorithms seek a solution that minimizes the inversion error over the given measurement-model pairs. There are many challenges that might prevent supervised machine learning algorithms from finding a proper mapping that can be generalized beyond the training dataset. One of the challenges is the lack of labeled data from a given survey area on which a model can be trained. For this reason, such algorithms must have a limited number of learnable parameters (i.e. shallow neural networks) and good regularization methods in order to prevent over-fitting and to be able to generalize well \cite[]{alfarraj2018petrophysical}.

Alternatively, a solution of the inverse problem can be sought in an unsupervised-learning scheme where the learning algorithm is given a set of measurements only $d$ and a forward model $\mathcal{F}$. The algorithm then learns by minimizing the following data misfit described by the following equation
\begin{equation}
    L(\Theta):=\mathcal{D}\left(\mathcal{F}\left(\mathcal{F}_{\Theta}^\dagger(d)\right),d\right)
    \label{eqn:unsupervised}
\end{equation}

Such formulation does not integrate well log data directly in the learning process. Furthermore, the forward model and its parameters must be chosen carefully to result in reasonable inversion. 

In this work, we proposed a \emph{semi-supervised machine-learning approach to seismic inversion that integrates both well log data misfit in addition to data misfit}. Semi-supervised learning enables the use of deep learning to seek better inversion without high data requirements as often required in supervised deep learning schemes. Formally, the loss function of the proposed workflow is written as 

\begin{equation}
    L(\Theta_1, \Theta_2):= \alpha\cdot \underset{\text{property loss}}{\underbrace{\mathcal{D}\left(\hat{m}, \mathcal{F}_{\Theta_1}^\dagger(d)\right)}} + \beta \cdot \underset{\text{seismic loss}}{\underbrace{ \mathcal{D}\left(\mathcal{F}_{\Theta_2}\left(\mathcal{F}_{\Theta_1}^\dagger(d)\right), d\right)}}
    \label{eqn:semi-supervised}
\end{equation}

where $\mathcal{F}^{\dagger}_{\Theta_1}$ is a learned inverse model parameterized by $\Theta_1$ and  $\mathcal{F}_{\Theta_2}$ is a learned forward model parameterized by $\Theta_2$. In addition, $\alpha, \beta \in \mathbb{R}$ are tuning parameters that govern the influence of each of the property loss and seismic loss, respectively.

\section{Methodology}
\begin{figure}[h]
    \centering
    \includegraphics[width=\linewidth]{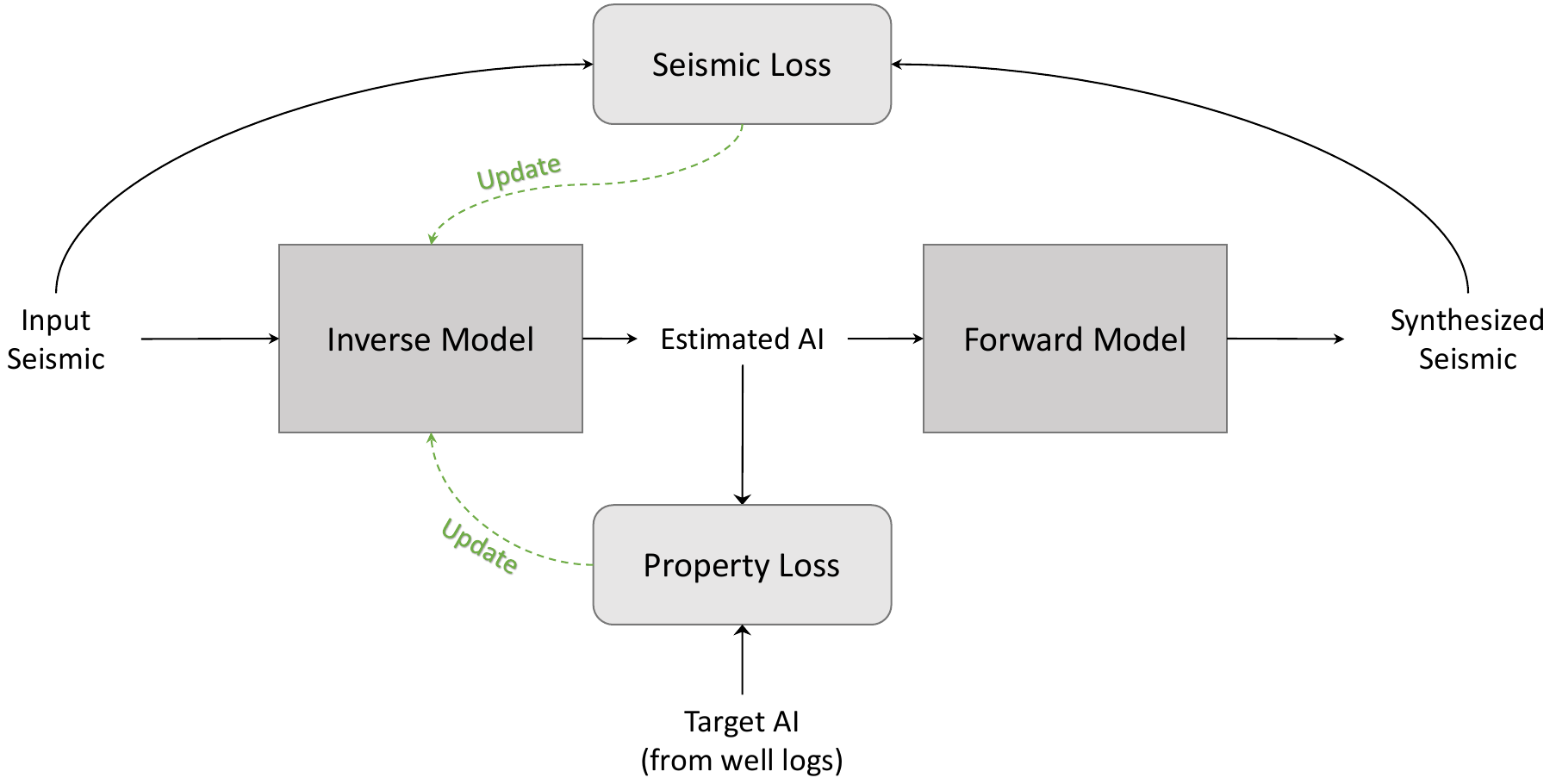}
    \caption{The proposed workflow.}
    \label{fig:workflow}
\end{figure}

The proposed workflow shown in Figure \ref{fig:workflow} consists of two main modules: the inverse model ($\mathcal{F}^\dagger_{\Theta_1}$) and a forward model ($\mathcal{F}_{\Theta_2}$); both of which have learnable parameters. The inverse model takes zero-offset seismic traces as inputs, and outputs the best estimate of the corresponding AI. Then, the forward model is used to synthesize seismograms from the estimated AI. The error (data misfit) is computed between the synthesized seismogram and the input seismic traces using the \textit{seismic loss} module for all traces in the survey. Furthermore, \textit{property loss} is computed between estimated and true AI on traces for which we have a true AI from well logs. The parameters of both the inverse model and forward model are adjusted by combining both losses as in equation \ref{eqn:system_setup}. 

In this work, we chose the distance measure ($\mathcal{D}$) as the Mean Squared Error (MSE). Hence, equation \ref{eqn:system_setup} reduces to: 
\begin{equation}
L(\Theta_1, \Theta_2) :=  \underset{\text{mean property loss}}{\underbrace{\frac{\alpha}{N_p}\lVert \hat{m} - \mathcal{F}^\dagger_{\Theta_1}(d)\rVert_2^2}} + \underset{\text{mean seismic loss}}{\underbrace{\frac{\beta}{N_s} \lVert d - \mathcal{F}_{\Theta_2}(\mathcal{F}^\dagger_{\Theta_1}(d)) \rVert_2^2}}
\label{eqn:loss}    
\end{equation}

where $N_\text{s}$ is the total number of seismic traces in the survey, and $N_p$ in the number of available well logs from which AI traces are obtained. In seismic surveys, $N_p \ll N_s$, therefore, the seismic loss is computed over many more traces that the property loss. On the other hand, the properly loss has access to direct high-resolution model parameters (well log data). To ensure stable learning, $\alpha$ and $\beta$ are chosen to balance learning from the two terms of the loss function. In this work, we chose $\alpha=0.2$, and $\beta=1$.

\subsection{Inverse Model}
\begin{figure*}
    \centering
    \includegraphics[width=\linewidth]{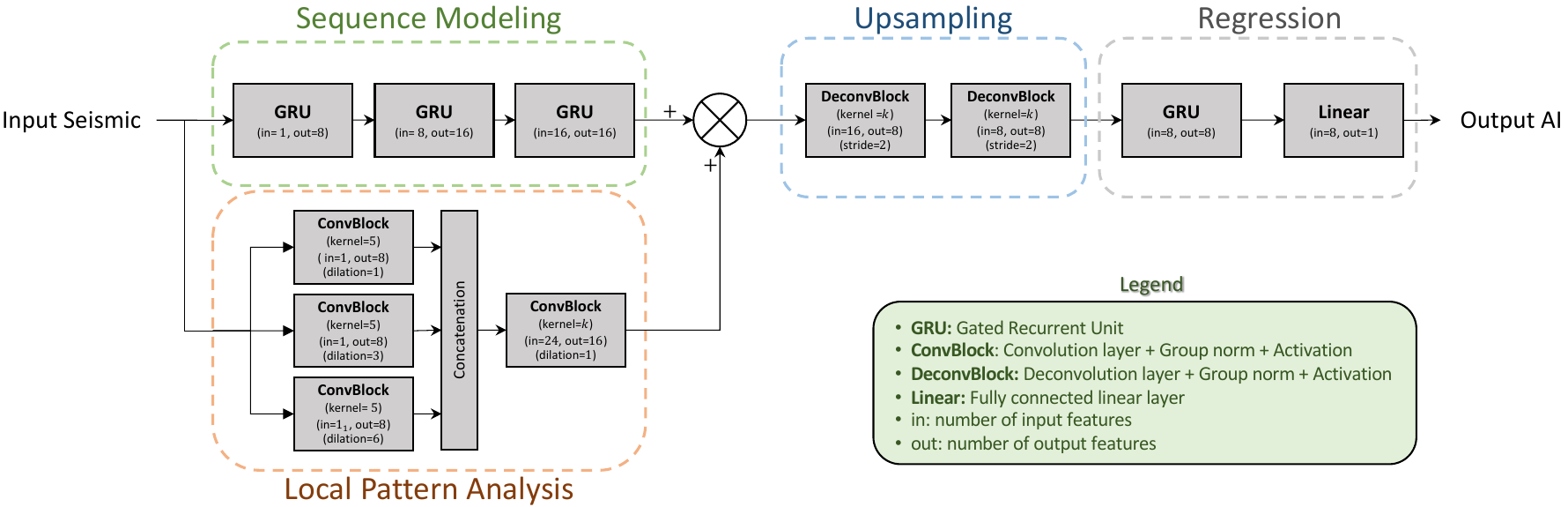}
    \caption{The architecture of the inverse model in the proposed workflow.}
    \label{fig:inverse}
\end{figure*}

The proposed inverse model in the proposed workflow consists of four main submodules (shown in Figure \ref{fig:inverse}). These submodules are labeled as \emph{sequence modeling}, \emph{local pattern analysis}, \emph{upsampling}, \emph{regression}. Each of the four submodules performs a different task in the overall inversion model.
% The \emph{sequence modeling} module models the temporal dynamics of seismic traces and produces features that best represent the low-frequency content of AI. The \emph{local pattern analysis} module extracts local attributes from seismic traces that best model high-frequency trends of AI trace. The {upsampling} modules take the sum of the features produced by the previous modules and upsamples them vertically. This module is added based on the assumption that seismic data are sampled (vertically) at a lower resolution than that of well log data. Finally, the \emph{regression module} maps the upsampled outputs from features domain to target domain (i.e., AI).  

\subsubsection{Sequence Modeling}
The \emph{sequence modeling} submodule consists of a series of Gated Recurrent Units (GRU) \cite[]{cho2014properties}. GRUs model their inputs as sequential data and compute temporal features based on the temporal variations of the input traces. In addition, they compute a state variable from future and past predictions that serve as a memory. The series of the three GRUs in the \emph{sequence modeling} submodule is equivalent to a 3-layer deep GRU. Deeper networks are able to model complex input-output relationships that shallow networks might not capture. Moreover, deep GRUs generally produce smooth outputs. Hence, the output of the \emph{sequence modeling} submodule is considered as the low-frequency trend of AI. 

\subsubsection{Local pattern analysis}
The \emph{local pattern analysis} submodule consists of a set of 1-dimensional convolutional blocks with different dilation factors in parallel. The output features of each of the parallel convolutional blocks are then combined using another convolutional block. Dilation refers to the spacing between convolution kernel points in the convolutional layers \cite[]{yu2015multi}. Multiple dilation factors of the kernel extract multiscale features by incorporating information from trace samples that are direct neighbors to a reference sample (i.e., the center sample), in addition to the samples that are further from it. A convolutional block (\textit{ConvBlock}) in Figure \ref{fig:inverse} consists of a convolutional layer followed by group normalization \cite[]{wu2018group} and an activation function. In this work, we chose hyperbolic tangent function as the activation function. 

Convolutional layers operate on small windows of the input trace. Therefore, they mostly capture high-frequency trends in traces. However, since convolutional layers do not have a state variable like recurrent layers, they do not capture low-frequency trends. Hence, the outputs of the \emph{local pattern analysis} and \emph{Sequence modeling} modules are added to obtain a full-band frequency content.  

\subsubsection{Upsampling}
The \emph{upsampling} submodule is used to compensate for the resolution mismatch between seismic data and well log data. Deconvolutional layers (also known as transposed convolutional or fractionally-strided convolutional layers) are upsampling modules with learnable kernel parameters unlike classical interpolation methods with fixed kernel parameters (e.g., linear interpolation). In addition, the stride controls the factor by which the inputs are upsampled. For example, a deconvolutional layer with a stride of ($s=2$) produces an output that has twice the number of the input samples (vertically). Deconvolutional layers have been used for various applications like semantic segmentation and seismic structure labeling \cite[]{noh2015learning,alaudah2018learning}.

A deconvolutional block (\textit{DeconvBlock}) in Figure \ref{fig:inverse} have a similar structure as the convolutional blocks introduces earlier. They are a series of deconvolutional layer followed by group normalization and an activation function. 

\subsubsection{Regression}
The final submodule in the inverse model is \emph{regression} which consists of a GRU followed by a linear mapping layer (fully-connected layer). The role of this module is to regress the extracted features from the other modules to the target domain (AI domain). The GRU in this module is a simple 1-layer GRU that augments the interpolated outputs by the \emph{upsampling} submodule using global temporal features. Finally, a linear affine transformation layer (fully-connected layer) takes the output features from the GRU and map them AI values.

\subsection{Forward Model}
The role of the forward model is to synthesize seismograms from AI. Forward modeling is commonly used in classical inversion approaches. However, in the work, we use a neural network to learn an appropriate forward model from the data. We used a simple 2-layer CNN to compute features from the AI traces, followed by a single convolutional layer that resembles a wavelet convolution in forward modeling. One of the advantages of using a learned froward model is that it automatically extracts the wavelet from the data.  

\section{Case Study}
In order to validate the proposed algorithm, we chose Marmousi 2 model \cite[]{martin2006marmousi2} (converted to time) as a case study. Marmousi 2 model is an extension of the original Marmousi synthetics model that has been used for numerous studies in geophysics for various applications including seismic inversion, seismic modeling, and seismic imaging. The model spans 17 km in width and 3.5 km in depth with a vertical resolution of 1.25 m.

\subsection{Training The Models}
To train the proposed inversion workflow, we chose $20$ evenly-spaced traces for training ($N_p=20$). For those training traces, we assume we have access to both AI and seismic data. For all remaining traces in the survey ($N_s=2721$ traces), we assume we have access to seismic data only.

First, the inverse and forward models are initialized with random parameters. Then, randomly chosen seismic traces in addition to the seismic traces for which we have AI traces in the training dataset are inputted to the inverse model to get a corresponding set of AI traces. The forward model is then used to synthesize seismics from the estimated AI. \emph{Seismic loss} is computed as the MSE between the synthesized seismic and the input seismic. \emph{Property loss} is computed as the MSE between the predicted AI and the true AI trace on the training traces only. The total loss is computed as a weighted sum of the two losses. Then, the gradients of the total loss are computed, and the parameters of the inverse model are updated accordingly. The process is repeated until convergence. 

\subsection{Results and Discussion}
\begin{figure}[ht!]
    \centering
    \subfigure[Estimated AI]{
    \includegraphics[width=\linewidth]{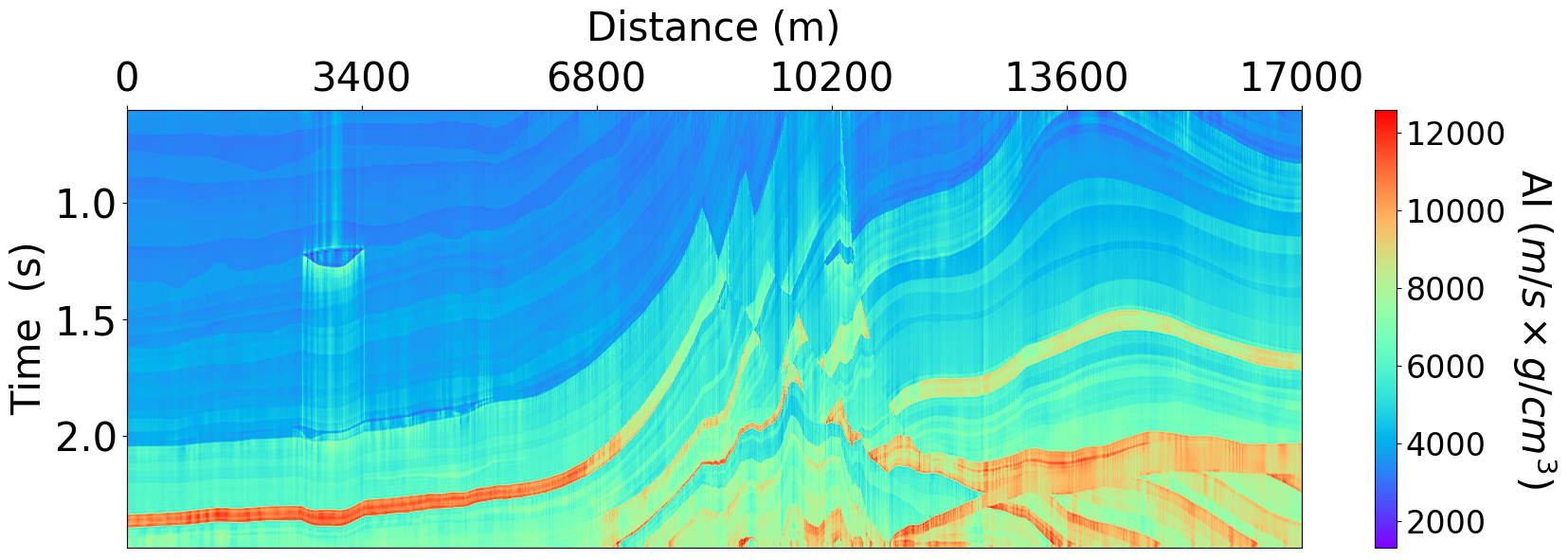}}
    
    \subfigure[True AI]{
    \includegraphics[width=\linewidth]{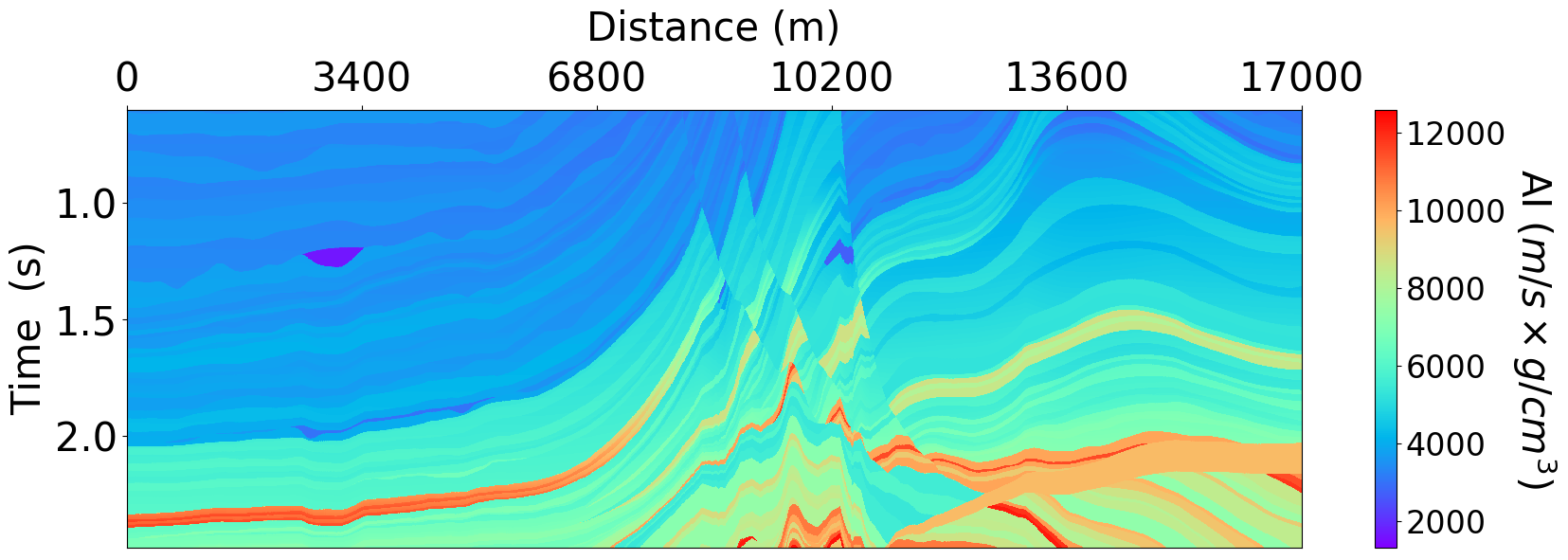}}
    \caption{Estimated AI and true AI for Marmousi 2 model.}
    \label{fig:results_section_AI}
\end{figure}

Figure \ref{fig:results_section_AI} shows estimated AI and true AI for the entire section. The shown predicted AI is the direct output of the inversion workflow with no post-processing. The jitter effect visible in the predicted AI is expected since the proposed workflow is based on 1-dimensional modeling with no explicit spatial constraints as often done in classical inversion methods. 

The traces around $x=3400$ m passe through an anomaly (Gas-charged sand channel) represented by an isolated and sudden transition in AI at $1.25$ ms. This anomaly causes the inverse model to incorrectly estimate AI. Since our workflow is based on bidirectional sequence modeling, we expect the error to propagate to nearby samples in both directions. However, the algorithm quickly recovers a good estimate for deeper and shallower samples of the trace. This quick recovery is mainly due to the reset-gate variable in the GRU that limits the propagation of such errors in sequential data estimation.

% \begin{figure}[ht!]
%     \centering
%     \includegraphics[width=0.9\linewidth]{Fig/raw/AI_traces.png}    
%     \caption{Selected estimated and true AI traces.}
%     \label{fig:results_traces}
% \end{figure}

\begin{figure}[ht!]
  \centering
    \includegraphics[width=0.9\linewidth]{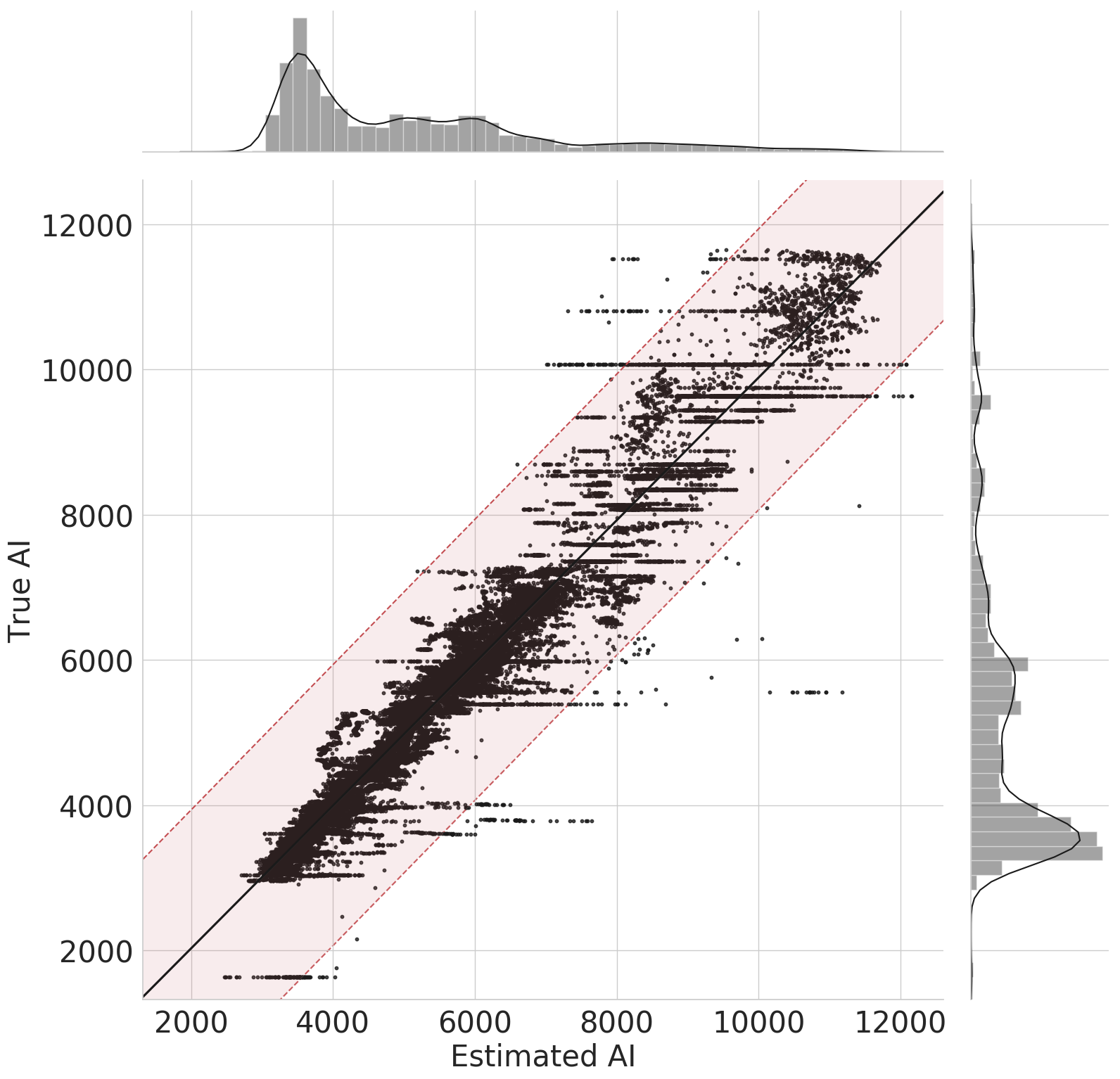}
    \caption{A scatter plot of the estimated and true AI. The shaded region includes all points that are within $\pm \sigma_{\text{AI}}$ of the true AI.}
    \label{fig:results_correlation}
\end{figure}

Furthermore, we show a scatter plot of the estimated and true AI in Figure \ref{fig:results_correlation}. The shaded region includes all points that are within one standard deviation of the true AI ($\sigma_{\text{AI}}$). The scatter plot shows a linear correlation between the estimated and true AI with the majority of the estimated samples within $\pm\sigma_{\text{AI}}$ from the true AI.

To evaluate the performance of the proposed workflow quantitatively, we use two metrics that are commonly used for regression analysis. Namely, Pearson correlation coefficient (PCC), and coefficient of determination ($r^2$). PCC is a measure of the linear correlation between the estimated and target traces. It is commonly used to measure the overall fit between the two traces. On the other hand, $r^2$ is a goodness-of-fit measure that takes into account the mean squared error between the two traces. The quantitative results are computed over the training traces and for all traces in the survey that were not used in the training (validation data). The results are summarized in Table \ref{tab:results}.

\begin{table}[ht!]
    \centering
    %EI
\begin{tabular}{c|cc}
    \hline
    \hline
    \diagbox{\textbf{Metric}}{\textbf{data}}& Training & Validation \\
    \hline
    \textbf{PCC}        &0.9836 &0.9809 \\
    $\boldsymbol{r}^2$  &0.9466 &0.9422 \\
    \hline
    \hline
\end{tabular}

    \caption{Quantitative evaluation of the estimated AI.}
    \label{tab:results}
\end{table}

The results in Table \ref{tab:results} shows that the performance of the proposed workflow on unseen data (validation) is very close to its performance on the training data, which indicates its generalizabilty beyond the training data. 

\section{Conclusion}
In this work, we proposed an innovative semi-supervised machine learning workflow for elastic impedance (AI) inversion from zero-offset seismic data. The proposed workflow was validated on the Marmousi 2 model. Although the training was carried out on a small number of AI traces for training, the proposed workflow was able to estimate AI for the entire Marmousi 2 model with an average correlation of $98\%$. The application of the proposed workflow is not limited to AI inversion; it can be easily extended to perform full elastic inversion as well as property estimation for reservoir characterization.

\section{Acknowledgements}
This work is supported by the Center for Energy and Geo Processing (CeGP) at Georgia Institute of Technology and King Fahd University of Petroleum and Minerals (KFUPM). 

\bibliographystyle{plain}
\bibliography{references}
\end{document}